\documentclass[12pt]{iopart}
\usepackage{epsf}
\usepackage{graphicx}
\def\gtsim {\lower .1ex\hbox{\rlap{\raise .6ex\hbox{\hskip .3ex
        {\ifmmode{\scriptscriptstyle >}\else
                {$\scriptscriptstyle >$}\fi}}}
        \kern -.4ex{\ifmmode{\scriptscriptstyle \sim}\else
                {$\scriptscriptstyle\sim$}\fi}}}

\def\ltsim {\lower .1ex\hbox{\rlap{\raise .8ex\hbox{\hskip .3ex
        {\ifmmode{\scriptscriptstyle <}\else
                {$\scriptscriptstyle <$}\fi}}}
        \kern -.4ex{\ifmmode{\scriptscriptstyle \sim}\else
                {$\scriptscriptstyle\sim$}\fi}}}

\def\aj{\rm{AJ}}                   
\def\araa{\rm{ARA\&A}}             
\def\apj{\rm{ApJ}}                 
\def\apjl{\rm{ApJ}}                
\def\apjs{\rm{ApJS}}               
\def\aap{\rm{A\&A}}                
\def\aapr{\rm{A\&A~Rev.}}          
\def\mnras{\rm{MNRAS}}             
\def\pasp{\rm{PASP}}               
\def\nat{\rm{Nature}}              

\begin{document}

\title{In Pursuit of the Least Luminous Galaxies}
\author{Beth Willman}
\address{Department of Astronomy, Haverford College, 370
  Lancaster Avenue, Haverford, PA, 19041}
\ead{bwillman@haverford.edu}

\begin{abstract}

  The dwarf galaxy companions to the Milky Way are unique cosmological
  laboratories.  With luminosities as low as $10^{-7}$ $L_{MW}$, they
  inhabit the lowest mass dark matter halos known to host stars and
  are presently the most direct tracers of the distribution, mass
  spectrum, and clustering scale of dark matter.  Their resolved
  stellar populations also facilitate detailed studies of their
  history and mass content.  To fully exploit this potential requires
  a well-defined census of virtually invisible galaxies to the
  faintest possible limits and to the largest possible distances.  I
  review the past and present impacts of survey astronomy on the
  census of Milky Way dwarf galaxy companions, and discuss the future
  of finding ultra-faint dwarf galaxies around the Milky Way and
  beyond in wide-field survey data.

\end{abstract}

\maketitle

\section{Introduction}

The least luminous known galaxies have historically been those closest
to the Milky Way.  Whether visually or with automated searches,
resolved stars reveal the presence of nearby dwarf galaxies with
surface brightnesses too low to be discovered by diffuse light
alone. Even until recently, nearly all cataloged dwarfs fainter than
$M_V = -11$ resided within the Local Group of galaxies (LG)
\cite{mateo98}.  In 1999 the LG contained 36 known members, of which
eleven are Milky Way (MW) satellites \cite{vandenbergh99}.  Four of
these eleven MW dwarf galaxies are less luminous than $M_V = -10$,
more than 10,000 times less luminous than the Milky Way itself.
Although such low luminosity dwarfs almost certainly contribute a
cosmologically insignificant amount to the luminosity budget of the
Universe, all eight of the Milky Way's
classical\footnotemark[1]\footnotetext[1]{``Classical'' will be used
  in the paper to refer to the Milky Way dwarf companions known prior
  to 2003.}  dwarf spheroidal companions ($ -9 > M_v > -13$, not
including Sagittarius or the Magellanic Clouds) have been studied in
extensive detail.  There is now a new class of ``ultra-faint'' dwarf
companions to the Milky Way known to have absolute magnitudes as low
as $M_V \sim -2$ (\cite{geha09a}, see Section 3). The resolved stellar
populations of these near-field cosmological laboratories have been
used to derive their star formation and chemical evolution histories
\cite{dolphin05} and to model their dark mass content in detail (see
article by L. Strigari in this volume and references therein).  These
complete histories of individual systems complement studies that rely
on high redshift observations to stitch together an average view of
the Universe's evolution with time.

The need for an automated, ``systematic, statistically complete, and
homogeneous search'' for LG dwarf galaxies has been known for some
time \cite{macgillivray87}.  A combination of theoretical results and
the advent of digital sky surveys have initiated a renaissance in the
pursuit of a well-measured sample of the least luminous galaxies.
This renaissance began in 1999, when simulations were used to
highlight the discrepancy between the number of dark matter halos
predicted to orbit the MW and the eleven observed to be lit up by
dwarf galaxies orbiting the MW \cite{klypin99,moore99}.  As the
resolution of simulations has increased over the last ten years, so
has the magnitude of this apparent discrepancy.  The most recent
simulations predict tens (M$_{halo}$ $> 10^6 M_{\odot}$, \cite{diemand08a}) or
even hundreds of thousands (M$_{halo}$ $> 10^5 M_{\odot}$, \cite{springel08b})
of dark matter halos around the Milky Way.  In light of this ``missing
satellite problem'', great attention has been paid to the total number
of Milky Way dwarf galaxies.  However, this is only one metric with
which to learn about the properties of dark matter.  The intrinsically
faintest dwarfs (which can only be found and studied close to the
Milky Way) likely inhabit the least massive dark matter halos that can host
stars.  Such dwarfs may thus provide the most direct measurement of the
mass spectrum, spatial distribution, and clustering scale of dark
matter.

What was initially viewed as a problem now provides an opportunity to
simultaneously learn about dark matter and galaxy formation
physics. Many studies have invoked simple models of galaxy formation
within low-mass dark matter halos to successfully resolve the apparent
satellite discrepancy within the context of $\Lambda$CDM
(e.g. \cite{bullock00a,benson02,kravtsov04,simon07a}).  See the review
article in this volume on ``Dark matter substructure and dwarf
galactic satellites'' by A. Kravtsov for more details on the original
missing satellite problem and on resolutions to this problem based on
models of star formation in low mass halos.

To untangle the extent to which dark matter physics, galaxy formation
physics, and incompleteness in the census of dwarf galaxies contribute
to this missing satellite ``opportunity'' requires a {\it
  well-defined} dwarf galaxy census that is as {\it uniform} as
possible to the {\it faintest limits}.  For example - {\it Well
  defined:} To compare observations of the MW dwarf population with
models requires a detailed, quantitative description of the current
census.  Quantitative assessments of the detectability of MW dwarfs in
recent survey data, plus an assumed spatial distribution of dwarfs,
enabled extrapolation of the known population to predict a total
number of $\sim$ 100 - 500 dwarf satellites
\cite{koposov08a,tollerud08a}. {\it Uniform:} Because the very least
luminous MW dwarfs ($M_V\sim -2$) can currently only be found within
50 kpc, it is presently unclear whether dwarfs can form with such
intrinsically low luminosities, or whether the tidal field of the
Milky Way has removed stars from these nearby objects. The epoch of
reionization and its effect on the formation of stars in low mass dark
matter halos also leaves an imprint on both the spatial distribution
\cite{willman04,busha09} and mass function of MW satellites
\cite{strigari07b,simon07a}.  Other studies have claimed that the
spatial distribution of MW satellites is inconsistent with that
expected in a Cold Dark Matter-dominated model
\cite{kroupa05a,metz08a}.  Robust tests of these models are not
possible without improving the uniformity of the MW census with
direction and with distance.  {\it Faintest limits:} Reaching the low
luminosity limit of galaxy formation is necessary to probe the
smallest possible scales of dark matter, the scales on which the model
faces the greatest challenges.  Moreover, a census to faint limits
over a large fraction of the MW's virial volume may yield enough
dwarfs to rule out dark matter models with reduced power on small
scales, although numerical effects presently inhibit concrete
predictions of such models \cite{wang07}.

The specific observational requirements to fully exploit the
population of MW dwarfs (and beyond) to effectively test dark matter
theories and/or to learn about galaxy formation therefore include:

\begin{itemize}
\item A census of dwarfs\footnote{We apply the term ``dwarf'' only to
    stellar systems that, through direct or indirect evidence, are
    known to be dark matter dominated either now or at any point in
    the past} that is minimally biased with respect to
  Galactic latitude, distance (at least out to the virial radius of
  the Milky Way), star formation history and structural parameters
\item A statistically significant sample of lowest luminosity
    dwarfs
  \item A sample of the least luminous dwarfs in a range of
    environments
\end{itemize}

This article focuses on the roles of wide-field, optical imaging
surveys of the past, present and future in the pursuit of a minimally
biased census of the least luminous galaxies.  In particular, it
focuses on automated analyses of resolved star counts as a method to
reveal these systems.  Since the visual searches of the 20th century,
new digital sky survey data has substantially progressed the
completeness and uniformity of the MW satellite census.  Although this
progress has already revolutionized the landscape of dwarf galaxy
cosmology, it has also revealed great incompleteness in our knowledge
of the least luminous galaxies.  Imminent and future surveys such as
the Southern Sky Survey \cite{keller07a}, PanSTARRS
1\footnote{http://pan-starrs.ifa.hawaii.edu/public/}, the Dark Energy
Survey \cite{DES}, and the Large Synoptic Survey Telescope
\cite{ivezic08} are poised to ultimately achieve the observational
requirements needed for MW dwarf galaxy cosmology.

\section{Discovering Milky Way dwarf galaxies, pre-SDSS}

All Milky Way dwarf galaxies known prior to 1990 were discovered in
visual inspections of photographic survey data.  Sculptor ($M_V =
-11.1$) and Fornax ($M_V = -13.1$) were discovered in 1938 by Shapley
\cite{shapley38a,shapley38b} in images obtained with a 24-inch
telescope at Harvard's Boyden Station. Leo I ($M_V = -11.9$), Leo II
($M_V = -10.1$), Ursa Minor ($M_V = -8.9$), and Draco ($M_V = -9.4$)
were discovered in the 1950's in the images obtained with a 48-inch
Schmidt telescope as part of the original Palomar Observatory Sky
Survey (POSS) \cite{harrington50,wilson55}.  The last Milky Way
companion discovered by an eyeball search was Carina (1977, $M_V =
-9.4$), found on photographic plates obtained in the Southern
hemisphere counterpart to the Palmar Observatory surveys - the ESO/SRC
Southern Sky Survey \cite{cannon77}.  Magnitudes listed above are from
\cite{grebel03}, except for Sculptor \cite{mateo98}.

At the time of Carina's discovery, it was hypothesized that ``The only
possibility for detecting new systems of this type would seem to be in
regions of relatively high foreground stars density, and will probably
require careful scanning under low power magnification or detailed
star counts'' \cite{cannon77}. This hypothesis was validated by the
discovery of Sextans in 1990 ($M_V$ = -9.5) \cite{irwin90} as an
overdensity of star counts from automated plate machine (APM) scans of
the same POSS and ESO/SRC survey data that had been carefully
inspected decades earlier. Sextans was discovered as part of the first
large-scale, automated search for Milky Way companions \cite{irwin94}.
The serendipitous discovery of the eleventh Milky Way companion,
Sagittarius, in 1994 \cite{ibata94} as a moving group of stars was the
final Milky Way dwarf discovered in the photographic survey data of
the 20th century.

Since the discoveries of the eleven classical Milky Way dwarf
satellites, Kleyna et al. \cite{kleyna97} and Whiting et
al. \cite{whiting07} conducted systematic searches of the COSMOS/UKST
survey of the southern sky and the POSS-II and ESO/SRC survey data,
respectively.  Whiting's eyeball, all-sky search resulted in the
discoveries of the Local Group dwarfs Antlia ($M_V$ = -11.2) and Cetus
($M_V$ = -11.3), but not new Milky Way satellites.  The closest
predecessor to the modern searches described in
Section~\ref{sec:sdss_searches}, Kleyna et al searched for
overdensities of resolved stars in spatially smoothed, pixellated maps
of star counts. Although their survey revealed no new dwarf galaxies,
they performed the first detailed characterization of the Milky Way
dwarf satellite census. 
The detection limits of these searches are discussed in
Section~\ref{ssec:limits_other}.


\section{Mining for the lowest luminosity dwarfs in the SDSS era}
\label{sec:sdss_searches}

Although the searches for dwarfs in the survey data available in the
20th century were impressively successful, empirical evidence
suggested that the census of Milky Way dwarf galaxies may not yet be
complete \cite{vandenbergh99,willman04}.  Since then, the Sloan
Digital Sky Survey (SDSS, \cite{york00}) revolutionized the field of
dwarf galaxy cosmology with the discoveries of 14 MW dwarfs (and
possible dwarfs) as overdensities of resolved stars: 2005 - Ursa Major
\cite{willman05a} and Willman 1 (originally known as SDSSJ1049+5103,
\cite{willman05b}); 2006 - Bo\"otes I \cite{belokurov06b}, Ursa Major
II \cite{zucker06a}, Canes Venatici I \cite{zucker06b}; 2007 - Segue
1, Coma Berenices, Leo IV, Canes Venatici II, Hercules (all announced
in \cite{belokurov07a}), Leo T \cite{irwin07a}, Bo\"otes II
\cite{walsh07a}; 2008 - Leo V \cite{belokurov08a}; 2009 - Segue 2
\cite{belokurov09a}.  Follow-up observations confirmed most of these
to be the most dark matter dominated (central M/L up to 1000
\cite{simon07a,geha09a}), least luminous ($-1.5 > M_V > -8.6$
\cite{martin08b}), and among the least chemically evolved galaxies
known in the Universe \cite{kirby08a,frebel09a}. Among these 14,
Willman 1, Segue 2, and Bo\"otes II have not yet been shown to be
dwarf galaxies rather than star clusters, or unbound remnants
thereof. The ultra-faint dwarfs are also predicted to be the most
detectable sources of gamma-rays from dark matter annihilation
\cite{martinez09,strigari08a}. In parallel with these Milky Way
discoveries, 11 new M31 satellite galaxies have been discovered,
primarily in large INT and CFHT surveys of M31 (And IX - And XX, $-6.3
> M_v > -9$
\cite{zucker04,martin06,majewski07a,zucker07a,ibata07,irwin08a,mcconnachie08a}).

The accomplishments of the SDSS dataset seems particularly remarkable
given that the data were obtained with 1 minute exposures taken on a
2.5m telescope, with a resulting $r$-magnitude limit of 22.2.  In
general, pushing
the census of resolved dwarf galaxies to lower luminosities and
greater distances can be accomplished by (1) obtaining photometry of
stars to fainter apparent magnitudes, and/or 2) more efficiently
suppressing the noise from point sources contaminating the signal from
stars belonging to a dwarf galaxy, and/or (3) reducing spurious
detections, the primary source of which had been cluster galaxies
mis-classified as point sources \cite{irwin94,kleyna97}.  The features
of the SDSS that facilitated (2) and (3) were its multi-band photometry
and accurate star-galaxy separation.  The digital camera and
uniformity of the survey also played key roles in its richness as a
hunting ground for dwarfs.

With a median luminosity of $M_V \sim -5$ ($10^4$ $L_{\odot}$), the
ultra-faints are up to ten million times less luminous than the Milky
Way.  All but Willman 1 and Leo T of the new Milky Way satellites are
invisible in the SDSS images, even in hindsight.  How was the presence
of these invisible galaxies revealed? The seventh data release
of SDSS, DR 7 \cite{dr7}, includes 11663 deg$^2$ of imaging and over
100 million cataloged stars. The searches that resulted in the
discoveries of the ultra-faint dwarfs were based only on analyses of
these cataloged stars.  The methods applied were all similar in
spirit, starting with the search of Willman et al. \cite{willman02a}.
The search technique summarized here is the specific method used in
the most recent automated search, that of Walsh, Willman \& Jerjen
(WWJ \cite{walsh09a}):

\begin{itemize}

\item {\bf Apply a color-magnitude filter to point sources:} The
  primary source of noise in searches for dwarfs in SDSS-depth data is
  MW stars. The middle left panel of Figure 1 shows that MW stars are
  smeared out in color and magnitude.  The red plume contains thin
  disk main sequence stars, the bright blue plume contains thick disk
  main sequence turnoff (MSTO) stars, and the faint blue plume
  contains halo MSTO and MS stars. However, the stars belonging to a
  dwarf galaxy will occupy a well-defined region of color-magnitude
  space.  All stars with colors and magnitudes inconsistent with a
  dwarf galaxy (at a particular distance) can thus be filtered out.
  WWJ used Girardi isochrones to define a color-magnitude (CM) filter
  for stars between 8 and 14 Gyr old and with -1.5 $<$ [Fe/H] $<$
  -2.3.  This filter is shown in the far left panel of Figure 1 for a
  dwarf galaxy with $d = 20$ kpc.  Unlike the matched filter technique
  of \cite{rockosi02}, stars outside of the filter are simply removed
  from the analysis.  No weighting is done, because the filter is not
  intended to exactly match stars from a specific stellar
  population. The CM filter was shifted to 16 values of $m-M$ between
  16.5 and 24.0 to search for dwarfs with $20 \ltsim$ d $\ltsim 600$
  kpc.  The middle left panel of Figure 1 shows that a 20 kpc
  color-magnitude filter contains substantial noise from both thick
  disk and halo stars.  The far right panel shows that a 100 kpc
  filter resides primarily between the two plumes and includes
  contamination from faint halo stars.  The horizontal branch (HB)
  extension of this 100 kpc filter passes through MSTO halo stars,
  suggesting that this HB extension may include more noise than signal
  from the least luminous systems.  Although the analysis of WWJ was
  automated and included no visual component, the result of this
  processing step is illustrated in the left and middle panels of
  Figure 2.  The Ursa Major I ultra-faint dwarf ($M_V$ = -5.5, $d =
  100$ kpc) is not visible in the star count map on the left.  After
  CM filtering, a slight overdensity of point sources becomes visible.

\begin{figure}[htbp]
\begin{center}
\includegraphics[height=6.5cm]{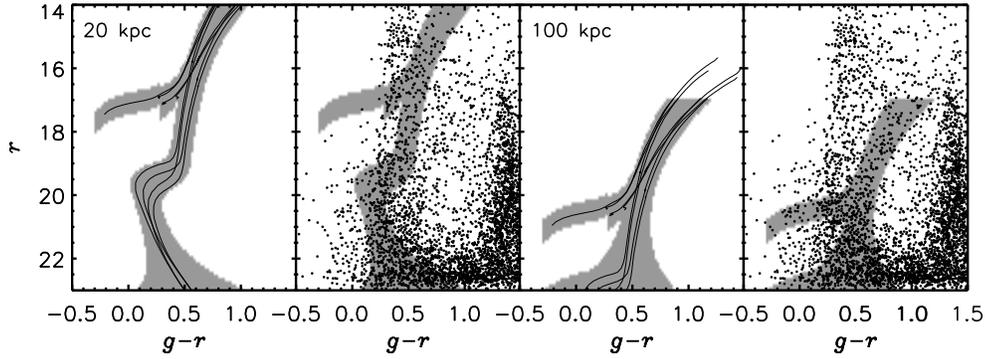}
\caption{A color-magnitude (CM) filter used to suppress the noise from
  foreground stars while preserving the signal from dwarf galaxy stars
  at a specific distance.  Far left and middle right: CM filters for
  an old and metal-poor stellar population at a distance modulus of
  16.5 and 20.0, respectively.  The solid lines show Girardi
  isochrones for 8 and 14 Gyr populations with [Fe/H] = -1.5 and -2.3.
  Middle left and far right: These CM filters overplotted on stars
  from a 1 deg$^2$ field to illustrate the character of the foreground
  contamination as a function of dwarf distance.}
\end{center}
\end{figure}

\item {\bf Create spatially smoothed image of stellar surface
    density:} As originally done in searches for nearby dwarf galaxies
  performed in the 1990's \cite{irwin94,kleyna97}, the number density
  map of stars passing CM filtering is smoothed with a spatial kernel
  to enhance the signals from resolved objects with angular scale
  sizes expected for nearby dwarf galaxies.  WWJ used only a 4.5$'$
  scale length filter, while \cite{koposov08a} applied filters of two
  different angular sizes. The result of this analysis step is
  illustrated in the right panel of Figure 2, which shows that Ursa
  Major I appears prominent in a spatially smoothed map of CM filtered
  stars.

\begin{figure}[htbp]
\begin{center}
\includegraphics[height=6cm]{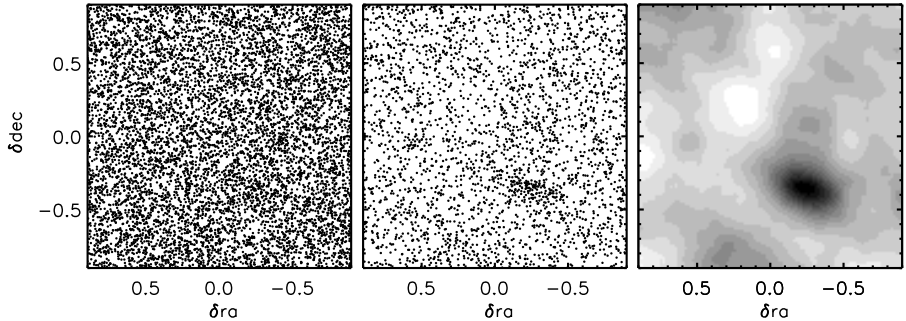}
\caption{Far left: Map of all stars in the field around the Ursa Major
  I dwarf satellite, $M_V = -5.5$, $d = $100 kpc. Middle: Map of stars
  passing the CM filter projected to $m-M = 20.0$ shown in the middle
  right panel of Figure 1. Far left: Spatially smoothed number density
  map of the stars in the middle panel.  The Ursa Major I
  dwarf galaxy has a $\mu_{V,0}$ of only 27.5 mag arcsec$^2$
  \cite{martin08a}.}
\end{center}
\end{figure}

\item {\bf Identify statistically significant overdensities:} A search
  of 10,000 deg$^2$ of SDSS data, optimized for dwarfs at 16 different
  distances, and a single choice of stellar population and scale size
  requires evaluating the statistical significance of 600 million data
  pixels that do not necessarily follow a Gaussian distribution of
  signal. Setting the detection threshold to select candidate dwarf
  galaxies was done by simulating numerous realizations of
  the search, assuming a random distribution of point sources, and
  permitting only one completely spurious detection.  The threshold
  is set to be a function of point source number density after CM
  filtering.

\item {\bf Follow-up candidates:} Regions detected above the detection
  threshold are considered candidates for MW dwarf galaxies. Although
  the threshold is set to prevent the detection of any stochastic
  fluctuations of a randomly distributed set of point sources
  \cite{walsh09a}, the detections are only ``candidates'' because
  resolved dwarf galaxies are not the only possible overdensities of
  point sources expected in the sky.  For example, fluctuations in the
  abundant tidal debris in the Milky Way's halo, or (un)bound star
  clusters could be detected.  Its essential to obtain follow-up
  photometry to find the color-magnitude sequence of stars expected
  for a dwarf galaxy, and also follow-up spectroscopy to measure the
  dark mass content (dark matter is required to be classified as a
  galaxy) based on the observed line-of-sight velocities.

\end{itemize}

This search algorithm is very efficient. In the WWJ search, the eleven
strongest detections of sources unclassified prior to SDSS were 11 of
the 14 (probable) ultra-faint Milky Way dwarfs.  All of these but
Bo\"otes II were known prior to the WWJ search.  See references in
Section 3 for details of the follow-up observations that confirmed
these objects to be dwarf galaxies.  Follow-up observations of as-yet
unclassified SDSS dwarf galaxy candidates are on-going by several
groups, including a group at the IoA at Cambridge (M. Walker, private
communication) and at the MPIA (N. Martin, private communication).
The {\it Stromlo Missing Satellites} team (PI H. Jerjen), is also now
obtaining and analyzing observations of the $\sim$ two dozen
candidates from the WWJ search of 9500 square degrees of SDSS DR6.

Because most probable candidates for dwarf galaxies have already been
followed up, it is possible that SDSS has already been completely
mined for ultra-faint dwarfs.  Nevertheless, it is essential to
concretely classify all objects identified down to the detection
threshold used to quantify the limits of a survey.  If there are
dwarf galaxies hiding in the low significance detections, then they
must be included when interpreting the properties of the global
population down to the observational limits.  If there are no dwarf
galaxies anywhere close to the detection thresholds, then there may
not be many unseen dwarfs with luminosities (distances) slightly
fainter than (a bit more distant than) those of similar dwarfs in the
known population.

\section{Current limitations of the census of Milky Way dwarfs}
\label{ssec:limits_sdss}
\label{ssec:limits_other}

As discussed in \S1, a well-defined census of dwarfs is essential to
use the MW dwarf galaxy population as a probe of dark matter and
galaxy formation physics. Astronomers have used a variety of
approaches to characterize the completeness of the Milky Way dwarf
census for more than 50 years, beginning with Wilson \cite{wilson55}
in 1955 who observed that ``The uniform coverage of the sky
  provided by the (Palomar Observatory) Sky Survey allows an
  estimate to be made of the probable total number of Sculptor-type
  galaxies in the local group.''

  Until this day, little is known about the possible population of MW
  dwarfs at $|b| < 20^{\circ}$ \cite{irwin94,kleyna97}, which includes
  1/3 of the volume around our galaxy, owing to obscuration by the
  Galaxy's disk.  A substantial fraction of the SDSS footprint is at
  $b > 30^{\circ}$, so no progress has yet been made on this severe
  observational bias at optical wavelengths.  Searches for satellites
  near the Galactic plane at radio and near-infrared wavelengths
  (2MASS) are less affected by disk obscuration than optical studies.
  Although two satellites have tentatively been discovered at these
  wavelengths (high-velocity cloud Complex H in HI survey data
  \cite{lockman03}, Canis Major in 2MASS \cite{martin04}), searches
  for MW dwarfs at non-optical wavelengths have not yet been very
  fruitful or quantified in detail.

  Likewise, the limitations of the Southern hemisphere dwarf galaxy
  census remain unchanged since the searches conducted with
  photographic plate data. Kleyna et al. \cite{kleyna97} derived
  detailed detection limits for their search by inserting simulated
  galaxies with the physical scale size of Sculptor into the COSMOS
  survey data.  They found that the Southern sky at $b <
  -15^{\circ}$ was complete to dwarfs closer than 180 kpc and as
  faint as 1/8 $L_{Sculptor}$, corresponding to $M_v = -8.8$.  Whiting
  et al. also quantitatively characterized the completeness of their
  visual search for dwarfs in the Southern Sky and estimated a
  limiting surface brightness ($25 < \mu_{lim} < 26$ mag
  arcsec$^{-2}$), with a 77\% completeness of dwarfs above this
  surface brightness limit \cite{whiting07}.

  It is thus likely that no dwarf similar to any of the 14
  ultra-faints discovered in SDSS data could have been found outside
  of the SDSS footprint. Within the SDSS footprint, the most extensive
  calculation of the limitations of the ultra-faint dwarf census is
  that of WWJ.  WWJ simulated the detectability of nearly 4 million
  artificial galaxies with a range of luminosity, scale size,
  distance, and Galactic latitude \cite{walsh09a}.  They estimate that
  the SDSS MW dwarf census is more than 99$\%$ complete within 300 kpc
  to dwarfs brighter than $M_V$ = -6.5 with scale sizes up to 1 kpc.
  Although this is a tremendous improvement, only four of the 14 new
  MW satellites are brighter than this limit.  $d_{90}$, the distance
  {\it at} which 90\% of dwarfs with some set of properties can be
  detected, is independent of the distribution of objects.  $d_{90}$
  is $\sim$ 35, 60 and 100 kpc for dwarfs with $M_V \sim$ -2, -3, and
  -4 with scale sizes similar to those of the known ultra-faints at
  like absolute magnitude.  (This is smaller than the distance {\it
    within} which 90\% of dwarfs with some set of properties can be
  detected.)  Larger scale length (lower surface brightness) systems
  are less detectable.  For example, systems with $M_v$ = -2 and a
  scale size of 100 pc or with $M_v$ = -4 and a scale size of 500 pc
  would have been undetectable in SDSS.  Koposov et
  al. \cite{koposov08a} derived quantitative detection limits for
  their SDSS search for ultra-faint dwarfs and found similar results.

The luminosity bias still present in the MW dwarf census as a function
of distance has several major implications.  First, the unknown
underlying radial distribution of MW dwarfs prevents assumption-free
predictions of their total number or luminosity function.  Second,
assumption-free comparisons between the observed and predicted spatial
distribution of MW dwarfs is still not possible.  However, studies of
the spatial distribution that only include the brighter MW dwarfs
($M_V < -5.5$) would provide initial insight into models.  Finally,
four of the MW ultra-faint companions (Willman 1, Bo\"otes II, Segue 1
and 2) have $L < 10^3 L_{\odot}$ ($M_V \gtsim -2.5$).  At present,
only $\sim$ 1/200 of the volume within the SDSS footprint has been
mined for such ultra-faints.  Are there pristine dwarfs in other
environments with such low luminosities?  Answering this question will
be critical for determining whether they have extremely low
luminosities because of nature (they formed that way) or nurture
(e.g. the tidal field of the Milky Way removed previously bound
stars).  Preliminary morphological studies suggest that the properties
of the nearest ultra-faints may have been affected by the MW's tidal
field.

These limitations and achievements do not substantively vary across
most of the SDSS footprint.  $\sim 50\%$ of the SDSS DR6 footprint
resides at $b > 50^{\circ}$ and only $\sim 10\%$ at $b < 30^{\circ}$.
$d_{90}$ is almost identical for dwarfs with $b = 53^{\circ}$ and $b =
73^{\circ}$, and was up to $\sim 25\%$ less for $b\sim 30^{\circ}$,
depending on $M_{V,dwarf}$.  The relatively weak variation with
latitude is owing to the CM filter (Figure 1) that does not include
stars with $g-r > 1.0$, cutting the majority of thin disk stars from
analysis.  Although the spatial variation is weak {\it on average},
regions of lower Galactic latitude plus longitude, or regions
containing substantial Sagittarius stream debris do have a lower
sensitivity for dwarfs.  For searches extending to $b \ltsim
30^{\circ}$, careful attention must be paid to the dependence of
detectability on Galactic direction.

\section{Mining for ultra-faint dwarfs post-SDSS}

To move from the excitement of discovery to more concrete comparisons
between observations and predictions will require progress on the
observational limitations described in \S4.  Here we highlight several
new and upcoming wide-field optical surveys that contain the qualities
necessary to make this progress.

The Southern Sky Survey (SSS) \cite{keller07a} and PanSTARRs (PS1) are
optical surveys of the entire Southern and Northern skies,
respectively.  The SSS is anticipated to begin survey operations at
the end of 2009, and PS1 has already begun obtaining survey data. The
SDSS filter set \cite{thuan76}, plus a Str\"omgren $u$ filter
will be used for the SSS, while SDSS $griz$ plus a $y$ filter at 1
micron is being used for PS1.  These surveys are both conducted on
small aperture telescopes (1.3m for SSS, 1.8m for PS1), with images of
the sky obtained repeatedly over a period of about 5 years.  The
co-added point source catalogs anticipated from these surveys will be
0.5 (SSS) to 1 (PS1) magnitude deeper than the SDSS catalog.

Searches for resolved dwarf galaxies in the SSS will be led by
H. Jerjen and the {\it Stromlo Missing Satellites} team and in PS1
will be lead by N. Martin at MPIA.  Between the SSS and PS1, a full
digital picture of the sky at optical wavelengths will be obtained,
nearly 75\% of it for the very first time. The region of sky at $b <
-20^{\circ}$ to be observed by the SSS should contain many
discoverable ultra-faint galaxies - perhaps a dozen by comparison with
those already known in the North.  These new surveys will also
substantially progress our understanding of the distribution of dwarfs
close to the disk.  However, mining for dwarfs at low $b$ will require
careful adjustments to the search techniques applied to SDSS data
owing to severe Galactic contamination and obscuration at low Galactic
latitudes.  For example, it has been common to use a $1^{\circ} \times
1^{\circ}$ running windows to measure the local density of the
foreground \cite{koposov08a,walsh09a}.  The steep spatial gradient in
the number density of disk stars at low $b$ will demand a more careful
characterization of the average point source counts when searching for
localized overdensities.

These imminent surveys will also reveal ultra-faint dwarfs throughout
a greater fraction of the Milky Way's virial volume.  A naive
extrapolation from the detectability of dwarfs in the SDSS yields
$\frac{d_{max,PS1}}{d_{max,SDSS}} = (\frac{f_{lim,PS1}}{f_{lim,SDSS}})^{0.5}$.
In this approximation, analyzing the PS1 star catalog with methods
analogous to those applied to SDSS data will reveal dwarfs (at $|b|
 <$ 20$^{\circ}$) to
distances $\sim $ 1.6 times farther, which is a factor of 4 in volume.
Despite this anticipated improvement, these surveys will not provide
an unbiased measurement of the ultra-faint dwarf galaxy population all
the way out to the virial radius of the Milky Way ($\sim$ 300 kpc).

Only a survey such as the planned Large Synoptic Survey Telescope
(LSST\footnote[4]{www.lsst.org}) project, currently scheduled to begin
survey operations in 2015, will potentially yield a measurement of the
ultra-faint dwarf galaxy population that truly satisfies all of the
observational requirements needed to fully exploit these objects for
dark matter and galaxy formation science.  LSST's primary mode will be
the planned ``deep-wide-fast'' survey that will observe 20,000 $deg^2$
of sky at $\delta < 34^{\circ}$ roughly 1000 times over 6 bands (SDSS
$ugriz$ plus $y$).  Single 15-second exposures have an anticipated
5$\sigma$ limit of $r = 24.5$ and the final 10-year co-added catalog
has an anticipated limit of $r = 27.5$ \cite{ivezic08}.

Using the same naive extrapolation of the detectability of dwarfs in
SDSS applied above to the PS1 survey, Tollerud et
al. \cite{tollerud08a} showed that an SDSS-like analysis of a 10-year
LSST-like catalog of stars would reveal $M_V = -2.0$ dwarfs to
distances of at least 400 kpc.  More luminous ultra-faints would be
detectable throughout the entire Local Group, and even beyond, based
on this sort of extrapolation.  Such a calculation assumes that the
number density of contaminating point sources passing color-magnitude
filtering (such as shown in Figure 1) does not substantially vary with
distance.  However, the landscape of the point source population at
magnitudes fainter than $r \sim 24$ does differ greatly from that in
the SDSS-depth data shown in Figure 1.

Figure 1 showed that thick disk and halo main sequence and main
sequence turnoff stars in the Milky Way were the primary noise in SDSS
searches.  At fainter apparent magnitudes, the number density of
unresolved galaxies, galaxies at high redshift that cannot be
distinguished from individual stars by morphology alone, rapidly
increases.  Figure 3 shows the $(V-I,V)$ color-magnitude diagram of
galaxies in the 9 arcmin$^2$ Hubble Ultra Deep Field (HUDF) with an
angular full-width half-max size smaller than 0.8$''$, the expected
average image quality of LSST.  Overplotted in red are the stellar
sources in the HUDF; they are outnumbered by galaxies by a factor of
75.

\begin{figure}[htbp]
\begin{center}
\includegraphics[height=8cm]{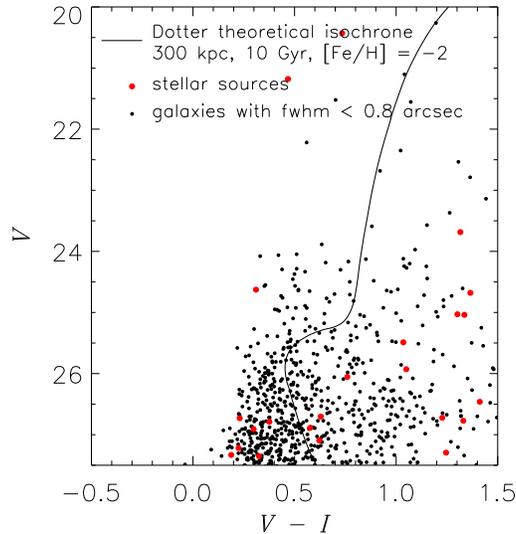}
\caption{Color-magnitude diagram of galaxies with small angular sizes
  and stellar sources in the Hubble Ultra Deep Field
  \cite{beckwith06}.  Galaxies outnumber stellar objects by a factor
  of 75 in this figure, suggesting that unresolved galaxies will be the
  primary source of contamination in searches for ultra-faint dwarfs
  in deep survey data.  Objects designated ``stellar'' in this image
  are those with $type$ $> 0.3$ in the HUDF catalog.}
\end{center}
\end{figure}

The CMDs in Figure 4 illustrate in more detail the point source
contamination expected in deep searches for resolved ultra-faint
dwarfs.  The left panel displays a
TRILEGAL\footnote[5]{http://stev.oapd.inaf.it/cgi-bin/trilegal}
\cite{girardi05} simulation of Milky Way stars in a one square deg
field at $(l,b)$ = (45,40).  The right panel displays a simulation of
the galaxy population as it will be observed by LSST.  The LSST image
simulation project (led by A. Connolly at UW) was based on a mock
catalog generated from the Millennium simulation
\cite{kitzbichler07}. The isochrone of an old and metal-poor stellar
population overplotted on each panel shows that red giant branch stars
belonging to a system $\sim$ 300 kpc away will be contaminated by MW
halo dwarf and subdwarf stars (the plume at $g-r \sim$ 1.0).  In
multi-color survey data of sufficient depth and photometric precision,
colors can be used to select stars based on temperature, metallicity,
and surface gravity \cite{lenz98}.  For example, it has been shown
that $g-r$ combined with $u-g$ separates metal-poor red giants at halo
distances from red dwarf stars in the disk of the Milky Way, but only
to $r\sim$ 17 in SDSS-depth data \cite{helmi03b}.  SDSS was not deep
enough in all filters to utilize photometric stellar classification to
distances beyond 25 kpc.  LSST will have small enough photometric
errors to photometrically select red giant stars at outer halo
distances.  Therefore, color-color selection of red giant stars at
outer halo distance may reveal both bound and unbound structure at MW
halo distances to unprecedentedly low surface brightnesses.

The overplotted isochrone also shows that the main sequence turnoff of
stars in an old and metal-poor stellar population in the MW's outer
halo will be severely contaminated by unresolved galaxies. The mock
galaxy catalog predicts $\sim$ 700,000 galaxies per deg$^2$ with $r
<$ 27.5 and $g-r <$ 1.5.  By contrast, the Trilegal model predicts
$\sim$ 35,000 stars per deg$^2$ with those same colors and magnitudes.
Based on the HUDF catalog, roughly half of the galaxies at the faint
magnitudes to be accessible by LSST have angular sizes smaller than
the expected median image quality of $0.8''$.  Unresolved galaxies
thus outnumber stars by a factor of 100 in observations down to $r =
27.5$ when only angular size is used to morphologically classify
objects, consistent with the results obtained from the small HUDF
field-of-view. 

The very least luminous ($M_V \gtsim$ -3) systems can only be
discovered by their MSTO and main sequence stars, because they have
few, if any, red giant branch stars.  The contamination by unresolved
galaxies could therefore be catastrophic for discoveries of such
systems at large distances, particularly because galaxies themselves
are clustered and thus do not provide a smooth background that can
easily be removed.  However, a combination of careful morphological
classification and color-color-magnitude filtering can be used to
drastically reduce the noise from unresolved galaxies.  In reality,
star-galaxy separation is not performed by a simple measurement of
angular size; the extended shapes of the light profiles of sources are
often used to discriminate between stars and galaxies.  For example
\cite{yee91} describes a method to use the curve-of-growth of the
light profile of individual objects to yield a morphological
star-galaxy classification.  This type of classification will still
yield a star catalog that is dominated by faint galaxies.  Galaxies
also have colors that differ from those of stars.  For example,
color-color information has been used to distinguish Milky Way stars
from unresolved galaxies at very faint magnitudes in the Deep Lens
Survey, a deep, ground based, survey in multiple optical filters
\cite{boeshaar03}.

\begin{figure}[htbp]
\begin{center}
\includegraphics[height=8cm]{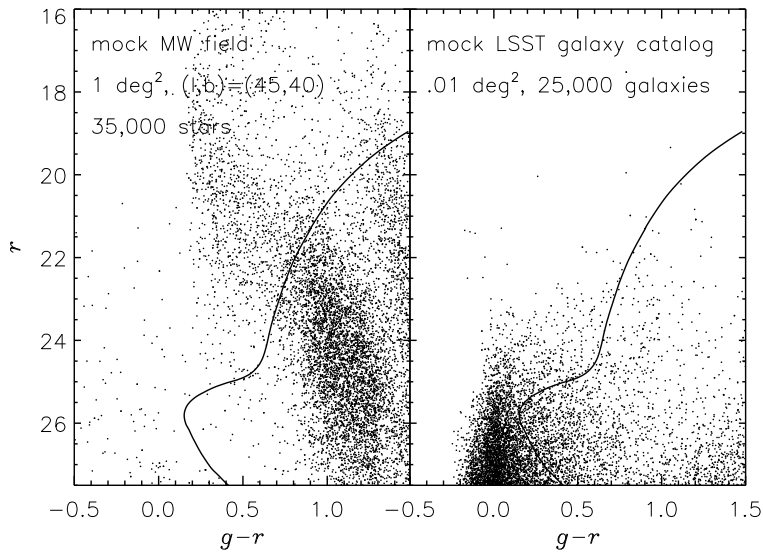}
\caption{Simulated observations of Milky Way stars and galaxies in an
  LSST-like survey.  Left panel: TRILEGAL simulated observation of
  Milky Way stars in a one deg$^2$ field at $(l,b)$ = (45,40).  Right
  panel: Simulated observation of galaxies in a 0.01 deg$^2$ field
  based on \cite{kitzbichler07} (A. Connolly, private
  communication). A Dotter isochrone for a 10 Gyr, [Fe/H] = -2.0
  stellar population at a distance of 300 kpc ($m-M \sim$ 22) is
  overplotted on both panels.  For clarity, only 1/4 of sources in
  each panel are plotted. LSST-like photometric uncertainties have not
  been added to the simulated data.}
\end{center}
\end{figure}

An important consideration for dwarf searches in LSST-depth data is
prospects for meaningful follow-up observations.  Follow-up imaging to
obtain deep CMDs has been needed to confirm many of the 14 known
ultra-faint dwarfs.  However, color-magnitude diagrams deeper than the
expected LSST limiting r-magnitude of 27.5 could likely not be
obtained from the ground.  Space-based follow-up to confirm new dwarfs
with JWST will probably also not be feasible, because the number of
dwarfs may be in the hundreds (with a higher number of candidates) and
because the fields-of-view of the cameras on JWST ($\sim$ 2.2$'$
$\times$ 2.2$'$) are smaller than the angular sizes expected for all
but the smallest scale size dwarfs.  With a half-degree field-of-view,
the camera on the Supernova Acceleration Probe (SNAP) could provide
the imaging needed to confirm the presence of relatively distant
dwarfs tentatively detected in LSST data.  There are not currently
plans for SNAP to be a pointed tool for such science.  Therefore the
number of resolved stars required for a certain ultra-faint detection
in very deep survey data will necessarily be higher than in SDSS-depth
data.  The spectroscopic resources now being used to measure the
masses of new ultra-faint objects (e.g. DEIMOS on Keck II, Hectochelle
on the MMT) are also already being pushed to their limits with the
dwarfs discovered in SDSS.  Much fainter or more distant dwarfs could
not be effectively studied with these resources, but instead will
require next generation 30m-class telescopes (such as a Giant Magellan
Telescope or Thirty Meter Telescope) and/or instrumentation.

A final consideration for searches based on resolved stars in an
LSST-depth data set - the possible crowding of stars belonging to more
distant satellites.  Although fewer stars are resolved in more distant
galaxies, the apparent angular separation of resolved stars decreases
with increasing distance.  If the average star separation is small relative
to the average full-width half-max of stars in the image, then an object
may be confusion limited and not identified in a standard
photometric pipeline.  Could ultra-faint dwarf galaxies become
confusion limited before they are, in theory, too distant to detect?
Using the Dotter stellar luminosity
functions\footnote[6]{http://stellar.dartmouth.edu/models/webtools.html}
\cite{dotter08} and assuming a star catalog as deep as the LSST
10-year coadd, the average spacing between resolved stars in a 10 Gyr,
[Fe/H] = -2.0 stellar population is roughly constant with distance for
100 kpc -- d$_{lim}$. $d_{lim}$ is the optimistic limiting detection
distance for dwarfs with $-2.5 > M_V > -7.5$. For ultra-faint Milky
Way satellites with scales sizes $\sim 50\%$ smaller (and thus smaller
angular separation between stars) than those of ultra-faints with
similar magnitudes, this average separation is expected to range
between 1$''$ and 2$''$.  Because this separation is larger than the
average image quality expected for LSST and because LSST will likely
reach its co-added depths by simultaneous photometering of numerous
exposures, rather than photometering a single stacked image, crowding
should not be a technical issue that will inhibit future dwarf
searches.

\section{Conclusion}

The next 15 years will be an exciting time for near-field dwarf galaxy
cosmology.  A lot hinges on the new class of ultra-faint galaxies that
was only discovered in the last 5 years, but that may be the most
numerous and cosmologically important class of galaxies.  However, to
effectively exploit these dwarfs as cosmological barometers will
require improvements on many observational limitations.  Several
wide-field, optical surveys are planned that may finally reveal the
true nature of the MW's satellite population and the true nature of
ultra-faint dwarfs.  Careful statistical analyses of star counts will
continue to be a primary method to identify ultra-faints, which are
known to have surface brightnesses as low as $\sim$ 27.5 mag
arcsec$^{-2}$.  Future surveys could possibly reveal such objects at
Mpc and greater distances by their diffuse light, rather than just by
individual stars.  Planned and current surveys at infrared wavelengths
will at minimum complement searches for dwarf galaxies done with
optical datasets and will provide important support for dwarf searches
near the Galactic plane.  The upcoming Vista Hemisphere Survey (PI
Richard McMahon) will image the entire Southern Sky in $J$ and $K_S$ 4
magnitudes deeper than 2MASS.  UKIDSS is in the middle of survey
operations and is obtaining 7000 deg$^2$ of IR imaging in the North to
a depth of K $\sim$ 18, including part of the Galactic plane.  These
surveys have the promise to open up enough new dwarf discovery space
to reveal systems not yet accessible in optical datasets.

Pointed surveys will also reveal low luminosity galaxies in other
systems, although they can not yet reveal objects as low luminosity as
many of the MW's ultra-faints.  Recently, \cite{chiboucas} identified
22 dwarf galaxy candidates as faint as $r$ = -10 around M81. They used
both eyeball evaluation and automated analysis of resolved stars in 65
square degrees of deep imaging.  The on-going PAndAS survey (PI
A. McConnachie) of 350 square degrees around M31 and M33 is expected
to reveal diffuse objects around these galaxies as faint as 32
magnitudes per square arcsecond.

The future will reveal whether we have yet seen the ultimate limit of
galaxy formation. The possibilities remain that either 1) the low
luminosities of the ultra-faint dwarfs are an artifact of nature,
rather than nurture, and/or 2) the present survey data are not deep
enough to reveal the very least luminous systems and a vast population
of ultra-faint dwarfs lie just beyond our fingertips.  Regardless, at
least dozens of ultra-faint satellites will be discovered in the near
future, with the possibility of hundreds or more.


\vskip 0.4cm

BW thanks both referees for comments that improved the clarity and the
quality of this paper.  BW also thanks Joe Cammisa at Haverford for
computing support and Marla Geha for inspiring Figure 2 of this
paper.

\vspace{2cm}


\end{document}